\documentclass[a4paper,preprint,aip,jcp,bibnotes,floatfix,11pt]{revtex4-1}
\usepackage{graphicx}

\usepackage[utf8]{inputenc}
\usepackage{graphicx}
\usepackage{tabularx}
\usepackage{epsfig}
\usepackage{subfigure}			
\usepackage[breaklinks=true]{hyperref}
\usepackage{lineno}
\usepackage[usenames,dvipsnames]{xcolor}
\usepackage{textcomp}
\usepackage{ulem}
\usepackage{url}
\usepackage{epstopdf}

\newcommand\grad{$^\circ$}

\begin{document}
 \title{Temperature-dependent templated growth of porphine thin films on the (111) facets of copper and silver}
\author{Katharina Diller}
\email[corresponding author: ]{katharina.diller@tum.de}
\affiliation{\mbox{Physik Department, E20, Technische Universität München, Garching, Germany}}
\affiliation{\mbox{Theoretische Chemie, Technische Universität München, Garching, Germany}}
\author{Florian Klappenberger}
\author{Francesco Allegretti}
\author{Anthoula C. Papageorgiou}
\author{Sybille Fischer}
\author{David A.\ Duncan}
\affiliation{\mbox{Physik Department, E20, Technische Universität München, Garching, Germany}}
\author{Reinhard J. Maurer}
\affiliation{\mbox{Theoretische Chemie, Technische Universität München, Garching, Germany}}
\author{Julian A.\ Lloyd}
\author{Seung Cheol Oh}
\affiliation{\mbox{Physik Department, E20, Technische Universität München, Garching, Germany}}
\author{Karsten Reuter}
\affiliation{\mbox{Theoretische Chemie, Technische Universität München, Garching, Germany}}
\author{Johannes V.\ Barth}
\affiliation{\mbox{Physik Department, E20, Technische Universität München, Garching, Germany}}


\begin{abstract}
The templated growth of the basic porphyrin unit, free-base porphine (\mbox{2H-P}), is characterized by means of X-ray photoelectron spectroscopy (XPS) and near-edge X-ray absorption fine-structure (NEXAFS) spectroscopy measurements and density functional theory (DFT).
The DFT simulations allow the deconvolution of the complex XPS and NEXAFS signatures into contributions originating from five inequivalent carbon atoms, which can be grouped into C-N and C-C bonded species.
Polarization-dependent NEXAFS measurements reveal an intriguing organizational behavior: 
On both Cu(111) and Ag(111), for coverages up to one monolayer, the molecules adsorb undeformed and parallel to the respective metal surface. Upon increasing the coverage, however, the orientation of the molecules in the thin films depends on the growth conditions. Multilayers deposited at low temperatures (LT) exhibit a similar average tilting angle (30\grad\ relative to the surface plane) on both substrates. 
Conversely, for multilayers grown at room temperature a markedly different scenario exists.
On Cu(111) the film thickness is self-limited to a coverage of approximately two layers, while on Ag(111) multilayers can be grown easily and, in contrast to the bulk \mbox{2H-P} crystal, the molecules are oriented perpendicular to the surface. This difference in molecular orientation results in a modified line-shape of the C~1\textit{s} XPS signatures that is dependent on the incident photon energy, which is explained by comparison with depth-resolved DFT calculations. Simulations of ionization energies for differently stacked molecules show no indication for a packing-induced modification of the multilayer XP spectra, thus indicating that the comparison of single molecule calculations to multilayer data is justified. 
\end{abstract}

\maketitle

\section{Introduction}
The combination of a well defined active site and a robust and flexible macrocycle, as well as the possibility to incorporate a multitude of metal centers and attach a wide range of substituents, make tetrapyrrole molecules, such as porphyrins and phthalocyanines, attractive for a wide range of applications. 
Moreover, porphyrins adsorbed on metal supports, whereby often coinage metal surfaces have been employed, offer an interesting playground to study how the interplay between molecule-molecule and molecule-substrate interactions influence the formation of nanostructures or organized layers. 
Consequently, porphyrins at interfaces have been extensively investigated both \textit{in vacuo} and in solution. 
One of the prominent tools of surface science, scanning tunneling microscopy (STM) is powerful especially for imaging single molecules or networks, but is typically less suitable for studying subtle details of molecular conformations or  molecular coverages beyond the first layer. This constraint, as well as the nature of certain nanotechnology applications\cite{Elemans2006, Koiry2010} such as chemosensors,\cite{Rakow2000, Filippini2006} data storage\cite{Lindsey2011} or nano-switches,\cite{Auwaerter2012} has resulted in the majority of the literature in this field to be focused on single-molecule or at most monolayer regimes for on-surface studies. In other fields, e.g., in catalysis, the dimensionality ranges from 0D (as single active sites\cite{Meunier1992,Mochida1983,Bhugun1996}) to 3D (in metal-organic frameworks\cite{Nakagaki2013,Goldberg2005}), whereas devices such as organic solar cells \cite{Campbell2004,Vilmercati2006} or organic light emitting diodes (OLEDs) \cite{Baldo1998} generally require comparatively thick layers.\cite{Sueyoshi2013,Murawski2013,Bruetting2013,Su2012,Hoppe2004} 
As the physicochemical properties of these systems often depend on the adsorption geometry and the orientation of the molecules both with respect to each other and the surface, it is of great importance to achieve excellent control over the growth processes of porphyrin films.

Various aspects regarding the formation of porphyrin mono- and multilayers can be addressed using complementary surface sensitive techniques. Low-energy electron diffraction (LEED) experiments, for example, provide information on the surface ordering and formation of superlattices.\cite{Krasnikov2010}
Photoelectron diffraction\cite{Duncan2010} and polarization-dependent near-edge X-ray absorption fine-structure (NEXAFS)\cite{Diller2012} provide insights into the bonding geometry and conformation;\cite{Auwaerter2007a} while X-ray photoelectron spectroscopy (XPS) casts light on the chemical environment of the molecules.\cite{Bai2009} The two latter X-ray spectroscopy techniques have the advantage that they can be employed for a wide range of coverages and do not require any ordering; therefore they are perfectly suited for the present investigations. 

In previously published work, studying the basic porphyrin unit free-base porphine (\mbox{2H-P}, Fig.\ \ref{fig:xps_stobe}a) on Ag(111)\cite{Bischoff2013} and Cu(111)\cite{Diller2013} substrates, a monolayer coverage was observed to behave in a broadly similar manner to larger, substituted porphyrins (e.g.,\ the ability to self-metalate on Cu and a planar adsorption geometry), however with remarkable differences (e.g.,\ island formation vs.\ repulsive interaction for 2H-P), which prompted the in-depth study, presented here, into the thin film growth modes on \mbox{2H-P} on these substrates.

We present an analysis of \mbox{2H-P} films of varying thicknesses deposited on both Cu(111) and Ag(111) substrates at room temperature (RT) and low temperature (LT) by means of XPS and NEXAFS. The first part of this study will focus on assigning the spectroscopic signatures of the XPS and NEXAFS measurements by comparing them to density functional theory (DFT) simulations. 
The main part of this work will focus on monitoring the temperature dependent \mbox{2H-P} thin film evolution, especially the molecular orientation within these films, using NEXAFS and XPS. The influence of the molecular stacking on the XPS signatures, showing that for porphine single molecule simulations are sufficient to reproduce the experimental multilayer data, will also be briefly analyzed. We will assess an observed photon energy dependence of the XP spectra through introducing depth-dependent DFT simulations. Hereby, we will address and fully characterize the influence of the choice of substrate and growth conditions on the resulting multilayer films.

\section{Experimental Details}
The XPS and NEXAFS data were recorded at the HE-SGM beamline of the BESSY II synchrotron radiation source in Berlin. The experiments were performed in an ultrahigh vacuum system with base pressures in the low 10$^{-10}$~ mbar (analysis chamber) and low 10$^{-9}$~ mbar (preparation chamber) regime. The Cu(111) and Ag(111) single crystals (Surface Preparation Laboratory, polished to $<$ 0.5\grad) were cleaned by repeated cycles of Ar$^{+}$ sputtering at 1~keV and subsequent annealing to 770~K (Cu) and 720~K (Ag), respectively. Prior to the experiments the \mbox{2H-P} powder (Livchem, purity 95\%) were degassed \textit{in vacuo} by heating to 433~K for several hours. The porphine layers were then deposited by organic molecular beam epitaxy from a quartz crucible held at 493~K onto the substrates which were either kept close to room temperature (RT, 300-320~K) or cooled down, with liquid N$_{2}$, to a temperature of 150~K (LT). 
The spectroscopic data were acquired for RT and LT with the sample held at the deposition temperature (unless otherwise stated in the text). We use the term monolayer (ML) for the maximum coverage obtainable with planar molecules adsorbed flat on the surface. The prefix (sub-) in Figs.\ \ref{fig:growth_cu} and \ref{fig:growth_ag} indicates that (i) the low-coverage signatures are similar to those of monolayers and (ii) that the presented data were acquired from samples with coverages of 1~ML or slightly less.

XP spectra were acquired with a hemispherical electron energy analyzer (VG Scienta R3000) in normal emission mode using a photon energy $\hbar\omega$ of 435~eV (unless otherwise stated) and a pass energy of 20~eV. After calibrating the binding energy scale against the Cu~3p$_{3/2}$ (75.1~eV)\cite{Thompson2009} or the Ag~3d$_{5/2}$ (368.3~eV)\cite{Thompson2009} line of the substrates, Shirley backgrounds were subtracted from the raw data.

All NEXAFS spectra were obtained in the partial electron yield (PEY) mode with a retarding voltage of \mbox{-150}~V. By rotating the sample with respect to the incoming beam the incidence angle $\theta$ between the $\vec{E}$-vector of the linear polarized light and the surface normal was changed. Spectra were recorded at $\theta=$ 25\grad, 53\grad\ (which is the magic angle for the given polarization of 90$\%$) and 90\grad. The photocurrent from a C-contaminated gold grid was measured simultaneously to the NEXAFS and the characteristic C K-edge (285.0 eV) was used to calibrate the NEXAFS photon energy.
After subtraction of a constant background signal and of the spectrum of a bare Cu(111) or Ag(111) crystal from the sample spectrum, the measured spectra were divided by a NEXAFS spectrum measured for a cleaned gold sample in total yield mode providing the transmission through the beamline. The edge jump was then normalized to one.\cite{Stoehr1992}

\section{Computational Details}
\subsection{StoBe}
The calculations for the XP and NEXAFS spectra in Figs.\ \ref{fig:xps_stobe} and \ref{fig:nexafs_stobe} were performed with the DFT program package StoBe\cite{StoBe2} using the revised Perdew, Burke and Ernzerhof exchange-correlation functional (RPBE).\cite{Hammer1999,Perdew1996} StoBe employs localized, Gaussian-type basis sets to describe the Kohn-Sham orbitals.

Prior to the calculation of the spectroscopic parameters, the geometry of the \mbox{2H-P} molecule was optimized using all-electron triple-zeta plus valence polarization type basis sets for the description of the nitrogen,\cite{Dunning1971} carbon\cite{Dunning1971} and hydrogen atoms.\cite{Huzinaga1965} 

For all calculations of the electronic structure the excitation center (carbon or nitrogen) is described by an IGLO-III\cite{Kutzelnigg1991} basis in order to improve the representation of relaxation effects in the inner shells. For all other atoms of the same element type effective core potentials (ECP)\cite{Pettersson1987} are applied to facilitate the identification of the core orbital of interest with negligible effects on the simulated spectrum.\cite{Pettersson1983}

Ionization (XPS) energies were obtained as
\begin{equation}
E_{\text{ion}} = E_{\text{tot}}(n_{1s} = 0) - E_{\text{tot}}(n_{1s} = 1)
\label{eq:xps_en}
\end{equation}
where $E_{\text{tot}}(n_{1s} = 1)$ and $E_{\text{tot}}(n_{1s} = 0)$ are the total energies of the ground state and the core hole state, respectively. The obtained energies are displayed as bars in Fig.~\ref{fig:xps_stobe} and Fig.\ S1 and broadened with Gaussian functions (the respective FWHM values are indicated in the figure captions) to obtain a continuous spectrum.  

For the simulation of the NEXAFS spectra the transition potential (TP) approximation was applied,\cite{Slater1972,Triguero1998} where the occupation of the 1$s$ orbital is set to 0.5. This allows the calculation of all final states (and therefore all possible transition energies) in one single SCF calculation. To improve the description of Rydberg and continuum final states additional large, diffuse [19s19p19d] basis sets\cite{Agren1997} were included (double basis set technique\cite{Agren1997,Triguero1998}). For better comparison with the experimental spectra the obtained discrete excitation energies and dipole transition matrix elements are broadened with Gaussians whose widths vary with energy according to  

$f (E)=\left\{\begin{array}{cl} \mbox{0.65~eV} , & \mbox{for }E\leq E_{\text{ion}}\\ \mbox{5.5 eV}, & \mbox{for } E> E_{\text{ion}}\mbox{+10~eV} \end{array}\right.$\\

for the C- K-edge in Fig.\ \ref{fig:nexafs_stobe}, with a linear increase from 0.65 to 5.5~eV in between. This broadening was empirically determined to reproduce the experimental spectra by accounting for the reduced lifetime of the $\sigma^{*}$ resonances which leads to increasing widths.\cite{Stoehr1992}

The missing core hole relaxation in the TP approximation can be corrected by shifting the spectra by the difference between $E_{\text{ion}}^{\text{TP}}$ and $E_{\text{ion}}$ from Eq.\ \ref{eq:xps_en}. Together with a relativistic correction of 0.1~eV\cite{Takahashi2004} the resulting total shift amounts to -1.4~eV. For a better comparison with the experiment an additional shift of 0.1~eV was applied in Fig.\ \ref{fig:nexafs_stobe}.

\subsection{FHI-aims}
In addition to the StoBe calculations, the numeric atom-centered orbitals (NAO) code FHI-aims\cite{Blum2009} was employed to simulate the XPS energies of stacked porphine units. The difference in bond lengths, compared to the StoBe geometry, of the optimized \mbox{2H-P} monomer are minimal ($\leq$ 0.01~\AA/ $\leq$ 1\%). The maximal deviation in bond lengths from experimental values determined by X-ray diffraction measurements in ref.\ \onlinecite{Webb1965} or the B3LYP-DFT calculations of ref.\ \onlinecite{Vyas2008} are around 2\%. 

In section \ref{subsec:charac} the influence of the stacking of \mbox{2H-P} molecules on the XP spectra is investigated. To this end, the geometries of four different dimers (see Fig.\ \ref{fig:dimers}) were optimized using PBE and tier2 basis sets. Additionally, to simulate long porphine chains periodic boundary conditions were applied, i.e., one \mbox{2H-P} molecule was placed in a supercell. The width $\Delta z$ of the supercell (and consequently the molecule-molecule distance $d$) as well as the geometry of the molecule were then relaxed to obtain the optimal $d$ for two different porphine chains (Fig.\ S3). The here employed exchange-correlation functionals do not properly account for dispersive forces which are crucial for the interaction of the molecules.\cite{MueckLichtenfeld2007} Therefore we additionally use the semi-empirical dispersion correction scheme of Tkatchenko and Scheffler.\cite{Tkatchenko2009} 

Ionization energies were calculated using equation \ref{eq:xps_en}. Tier 2 and tier 3 basis set calculations using PBE, as well as B3LYP\cite{Becke1993a} calculations on a tier 2 light level resulted in the same monomer spectra as were obtained with StoBe. To avoid charging problems using periodic boundary conditions, we did not calculate XP spectra of the infinite chains, but cut a \mbox{2H-P} trimer from the optimized chain and simulated the spectra of the central molecule, under the assumption that the influence of the next-nearest neighbors on the core level spectra is negligible.

\section{Results and Discussion}

\subsection{Spectroscopic characterization of 2H-P}
\label{subsec:charac}
Peak assignment is crucial to the interpretation of spectroscopic data, i.e., determining which peak originates from which part of the molecule is necessary in order to gain insight from the measurement. Of special interest, when studying comparatively complicated molecules like porphyrins, is the analysis of basic units such as the free-base porphine. Understanding such basic units in these molecules can (according to the building block principle\cite{Stoehr1992}) allow the deconvolution of the spectra of these species, assuming the subunits are weakly interacting. This section is therefore dedicated to a detailed analysis of the XPS and NEXAFS signatures of free-base porphine by comparison to theoretical calculations.

\begin{figure}[p]
 \includegraphics[width=0.45\columnwidth,clip]{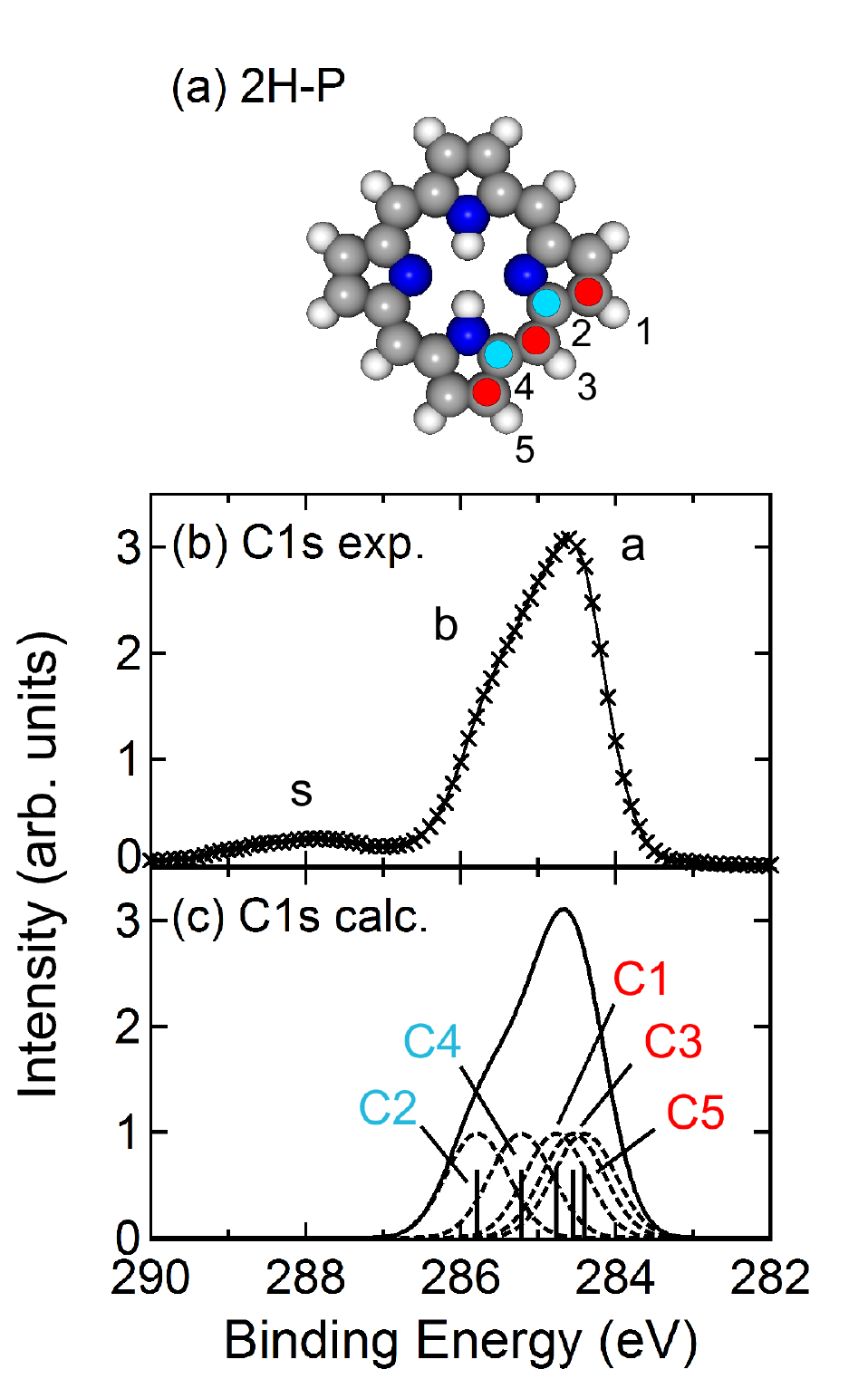}
 \caption{C~1\textit{s} XPS signature of porphine. (a) A schematic model of free-base porphine (2H-P) (C: gray, N: blue, H: white spheres). The numbering refers to the five types of inequivalent carbon atoms present in the molecule. The (b) experimental C~1\textit{s} XPS signature of porphine, resulting from a multilayer on Cu(111), which is well reproduced by (c) the simulated spectrum (broadening: 0.95~eV).}
\label{fig:xps_stobe}
\end{figure}

All free-base porphyrin molecules comprise four nitrogen atoms that form two, chemically inequivalent species: iminic (N) and pyrrolic (NH) (cf.\ Fig.\ \ref{fig:xps_stobe}a). Hence, N~1\textit{s} XP spectra of free base porphyrin molecules, including 2H-P,\cite{Krasnikov2008,Diller2013} typically feature two peaks that are separated by approximately 2~eV.\cite{Niwa1974,Yamashige2005a} 
This energy separation is reduced when the nitrogen atoms strongly interact with a substrate, such as the case with Cu(111),\cite{Diller2012,Buchner2011a} while it is preserved on more weakly interacting surfaces\cite{Buchner2011a} and in multilayers.\cite{DiSanto2011} 
The nitrogen region is of special interest for following chemical reactions such as the metalation of free-base porphyrins\cite{Gottfried2006, Doyle2011,Nowakowski2013} and is therefore frequently addressed.
For this reason the following discussion focuses on the less analyzed carbon region of \mbox{2H-P} (see supporting information for the corresponding nitrogen data). 
\begin{figure}[p]
 \includegraphics[width=0.4\columnwidth,clip]{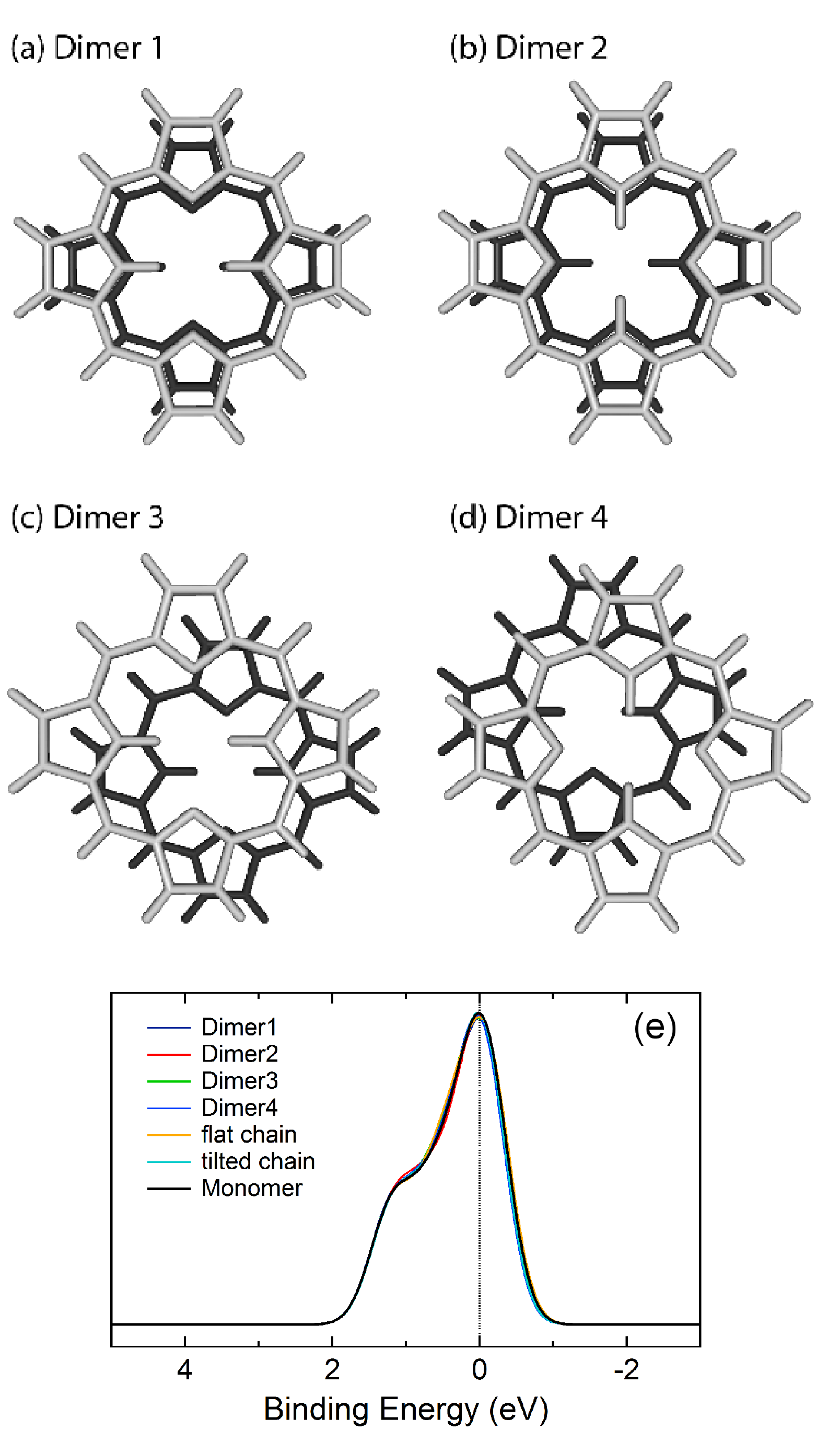}
 \caption{Influence of the porphine dimer geometries on the C~1\textit{s} XP spectrum: The simulated photoemission spectra of four differently stacked \mbox{2H-P} dimers (a-d, broadening: 0.75~eV) exhibit only negligible deviations from that of the monomer (Fig.\ \ref{fig:xps_stobe}).}
\label{fig:dimers}
\end{figure}
The C~1\textit{s} region of a \mbox{2H-P} multilayer grown on Cu(111) at LT (see Methods and Section \ref{subsec:multi_cu}) exhibits three main structures (Fig.\ \ref{fig:xps_stobe}b): a dominant peak at 284.6~eV (a), a shoulder at a slightly higher binding energy (b), and a broad feature at 288.0~eV (s). 
Features a and b are well reproduced by the simulated spectra of the five chemically non-equivalent carbon atoms in 2H-P (Fig. \ref{fig:xps_stobe}) allowing a clear assignment of these features. The contributions to these two features can be grouped according to whether the respective carbon atoms are directly bound to a nitrogen atom (C2/C4, shoulder b) or to other carbon atoms (C1/C3/C5,  feature a). It is important to note that feature s is not reproduced. This failure can be reconciled by considering its position and shape, which is a clear indication of a shake-up peak; an electron-electron interaction that cannot be modeled by employing current Density Functional Approximations.
\begin{figure}[p]
 \includegraphics[width=0.45\columnwidth,clip]{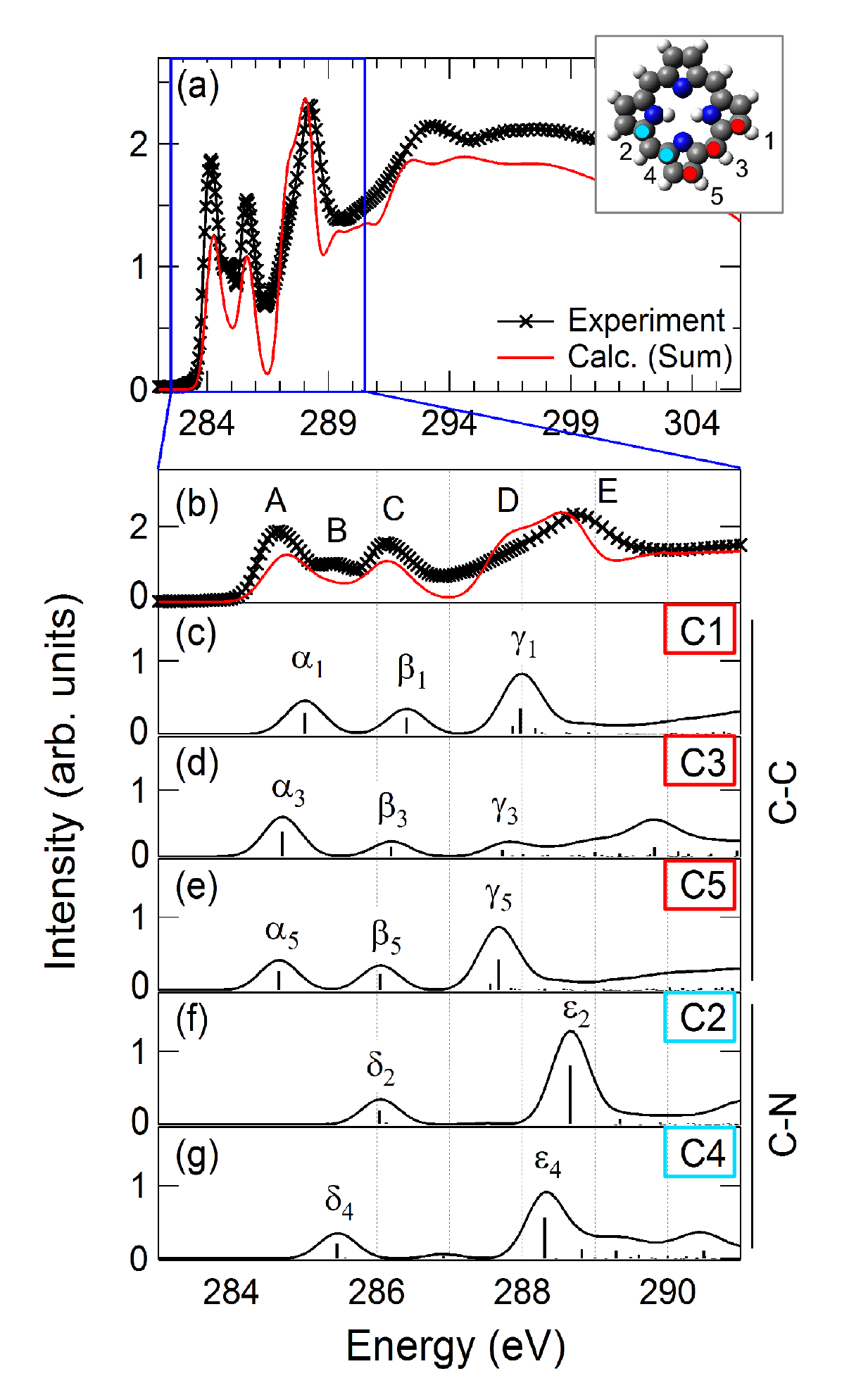}
 \caption{(a) The experimental NEXAFS C K-edge signature of 2H-P (black markers) measured with an incidence angle of 53\grad (i.e.\ the ``magic angle``) of a 2H-P multilayer on Cu(111) compared against the simulated spectrum (red continuous line). (b) A zoom-in of the  $\pi^*$ region and (c-g) the corresponding theoretical curves originating from the five inequivalent carbon atoms are also shown, indicating that the spectrum can be deconvoluted in two main contributions: C-C and \mbox{C-N} bonded carbon atoms. Initial state effects (consistent with XPS) are reflected in the shift of the respective curves in each group.}
\label{fig:nexafs_stobe}
\end{figure}

\begin{figure*}[p]
 \includegraphics[width=0.85\textwidth,clip]{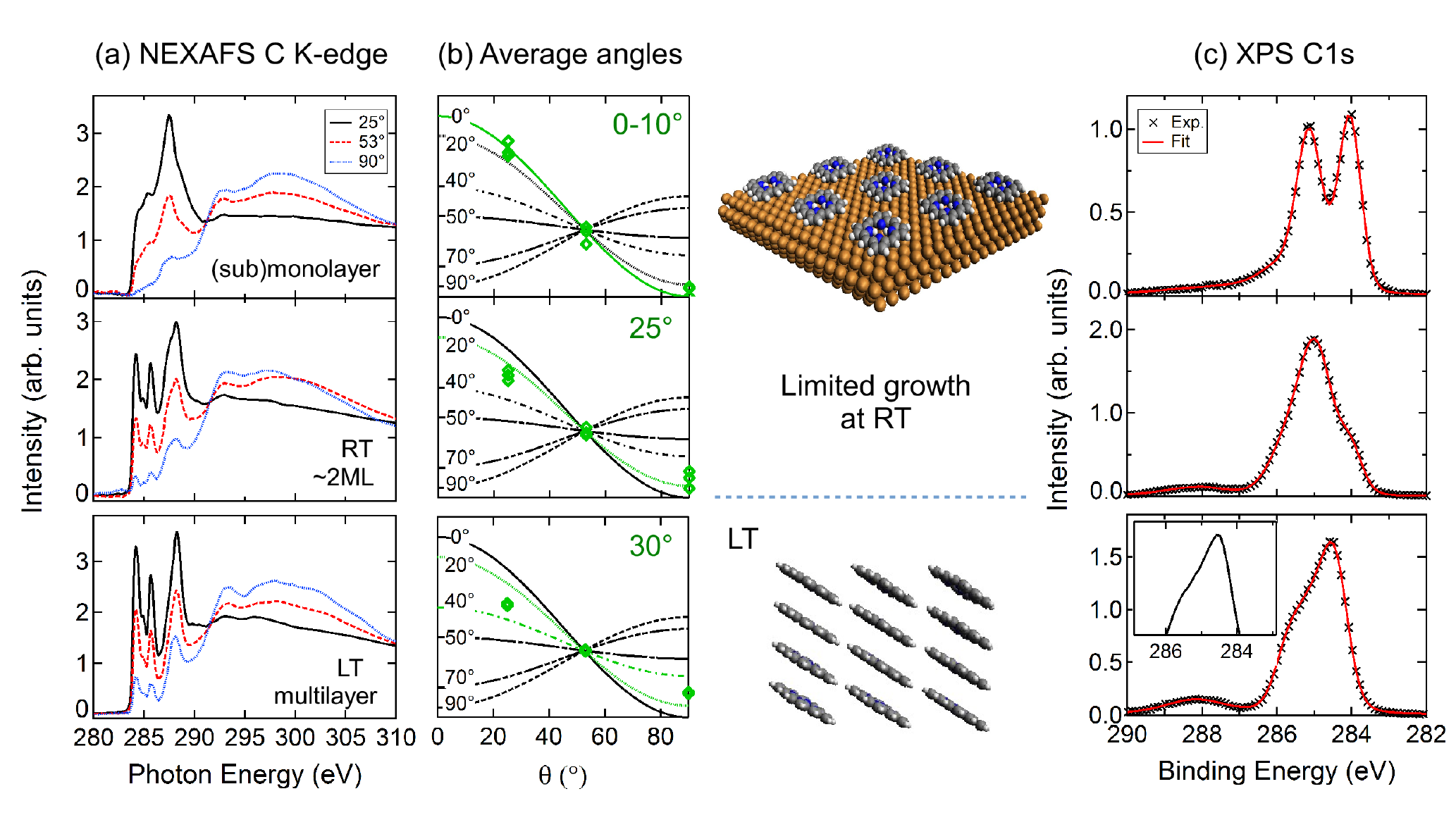}
 \caption{Growth of \mbox{2H-P} on Cu(111) at room and low temperature, monitored by (a) polarization-dependent C K-edge NEXAFS showing a (b) variation in the average tilt angle of the molecule. First layer molecules (top panels) interact strongly with the copper surface (see also ref.\ \onlinecite{Diller2013}) as evidenced by the distinct change in the (a) NEXAFS and (c) XPS signatures. At room temperature the growth is limited to approximately two layers (middle row panels). Conversely, for lower substrate temperatures it is possible to obtain multilayers with an average adsorption angle of 30\grad\ (bottom row panels). The shape of the corresponding C~1\textit{s} spectrum (c, bottom, inset) differs from that of the RT-phase on Ag(111) (Fig.~\ref{fig:growth_ag}c).}
\label{fig:growth_cu}
\end{figure*}

To study the influence of different stacking modes on the photoemission spectra, the geometries of four different porphine dimers (Fig.\ \ref{fig:dimers}) were optimized. In agreement with the results described in ref.\ \onlinecite{MueckLichtenfeld2007}, the symmetric configurations (Dimer 1 and Dimer 2) were found to be less energetically favorable, as indicated by the binding energies (Table SVI) and and by the fact that, for small deviations from the ideal symmetric configuration, the geometries always converged to Dimers 3/4. Likewise, the tilted chain is more stable than the symmetric one (Fig.\ S3). The simulation of the respective ionization energies shows no indication for a packing-induced modification of the XP spectra (Fig.\ \ref{fig:dimers}e), thus indicating that the comparison of single molecule calculations to multilayer data is justified. 

By the same methodology, the NEXAFS K-edge spectra of the nitrogen and the carbon regions (Fig.\ \ref{fig:nexafs_stobe}) can be analyzed. The N K-edge spectrum of \mbox{2H-P} (see supporting information and ref.\ \onlinecite{Diller2013}) is characteristic for porphyrins\cite{Goldoni2012,Narioka1995} and can be deconvoluted into two sets of spectra, originating from the two chemically different nitrogen species.\cite{Schmidt2010,Polzonetti2004,Diller2012} The great similarity to the spectra of substituted porphyrins, such as \mbox{2H-TPP}, indicates that the electronic structure of the macrocycle, or at least its center, is hardly affected by the attached substituents as long as they are neither of strong electron accepting nor donating character. 

Fig.\ \ref{fig:nexafs_stobe}a compares the measured C K-edge multilayer data (black markers) to the simulated curve (red continuous line). The analysis presented here specifically focuses on the $\pi^*$ region (Fig.\ \ref{fig:nexafs_stobe}b) which exhibits five main features from 284~eV to 291~eV. The five single, simulated spectra originating from the chemically different carbon atoms in the molecule (Fig.\ \ref{fig:nexafs_stobe}c-g) can be grouped according to the number and position of their dominant transitions: 
While the C-C bonded species (C1, C3, C5; cf.\ Fig.\ \ref{fig:xps_stobe}a) give rise to three main contributions ($\alpha$, $\beta$, $\gamma$), only two dominant resonances ($\delta$, $\varepsilon$) are present in the curves of the nitrogen bonded carbon species (C2, C4). The $\alpha$ and $\delta$ resonances correspond to transitions to the LUMOs of the transition potential state and their positions qualitatively and quantitatively follow the XPS binding energies, indicating that the respective shifts of the spectra are predominantly an initial state effect. The position of the experimental peaks A-E and the computed resonances $\alpha$-$\epsilon$, as well as a deconvolution of the measured curves, can be found in Tables SII and SIII (Supporting Information).
The deconvolution of the experimental spectrum provides an explanation for the changes in the C K-edge of \mbox{2H-P} on Cu(111) after self-metalation to \mbox{Cu-P} as described in ref.\ \onlinecite{Diller2013}. Peak A, which consists of C-C bonded carbon contributions, remains almost unchanged, while peak B (mainly due to transition $\delta_4$) vanishes, as after the metalation iminic nitrogen species are no longer present in the molecule. In a similar way, the multilayer/DFT characterization can be used to draw conclusions on the molecule-substrate interaction in porphyrin monolayers (refs.\ \onlinecite{Bischoff2013,Diller2013} and sections \ref{subsec:multi_cu} and \ref{subsec:multi_ag}).

\subsection{Temperature-dependent growth on Cu(111)}
\label{subsec:multi_cu}

\begin{figure*}[p]
 \includegraphics[width=0.85\textwidth,clip]{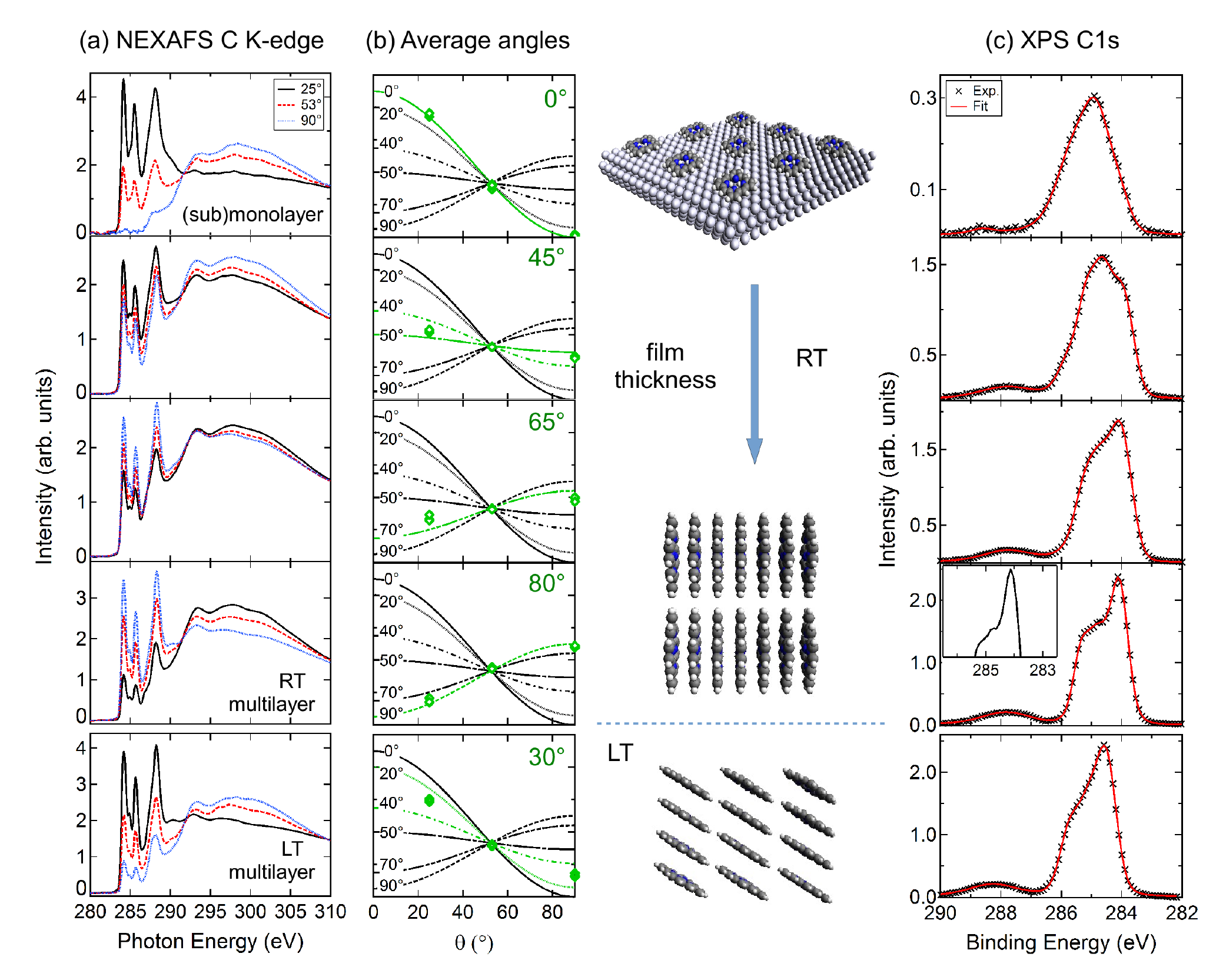}
\caption{Growth of 2H-P on Ag(111) at room and low temperature, monitored by (a) polarization-dependent C K-edge NEXAFS showing a (b) variation in the average tilt angle of the molecule. (c) The corresponding C~1\textit{s} XPS curves (markers: data points, red line: fit as guide to the eye) show the transition from a (sub)monolayer signature to the multilayer one (cf.\ Figs. \ref{fig:xps_stobe} and \ref{fig:lambda}). For LT-grown multilayers the average angle is strongly decreased (bottom).}
\label{fig:growth_ag}
\end{figure*}

Previously we have reported on the adsorption of (sub)monolayers of \mbox{2H-P} on Cu(111)\cite{Diller2013} and Ag(111),\cite{Bischoff2013} studied with STM, XPS, and NEXAFS. Polarization-dependent NEXAFS is an efficient tool to obtain information on the conformation and orientation of adsorbed molecules, as the intensities of the NEXAFS features depend on the photon incidence angle, $\theta$, i.e., the angle between the linear polarization of the light and the surface normal. For aromatic systems such as the porphine, the $\pi^*$ states are derived from $p_z$ orbitals oriented perpendicular to the molecular plane. If the aromatic $\pi^*$  system lies parallel to the surface the intensity of the corresponding transitions in the spectra exhibits a maximum for $\theta$ = 0 and vanishes for $\theta$ = 90\grad.\cite{Stoehr1992} Hence, the polarization-dependent NEXAFS curves (Figs.\ \ref{fig:growth_cu}a and \ref{fig:growth_ag}a) are fitted with Gaussian line shapes (see Fig.\ S4 for an exemplary fit) and the obtained relative intensities are compared to theoretical expected curves (Figs.\ \ref{fig:growth_cu}b and \ref{fig:growth_ag}b) to determine the adsorption angle between the aromatic system and the substrate.

For coverages up to 1~ML on both Cu(111) (Fig.\ \ref{fig:growth_cu}a, top) and Ag(111) (Fig.\ \ref{fig:growth_ag}a, top) \mbox{2H-P} adsorbs with its molecular plane parallel to the surface, as indicated by the vanishing 90\grad-curve (blue dotted line). On Ag(111) the molecules are clearly planar,\cite{Bischoff2013} while a small deformation (10-15\grad) cannot be ruled out on Cu(111), as the stronger interaction with the substrate leads to a general broadening of the NEXAFS features.\cite{Diller2013} The partially quenched first peak was attributed to a partial filling of the LUMO due to electron transfer from the substrate, in agreement with calculations for \mbox{2H-P} on Cu(110).\cite{Dyer2011} The strong interaction is also reflected in the drastically modified appearance of the C~1\textit{s} XP spectrum (Fig.\ \ref{fig:growth_cu}c, top), compared to the simulated gas phase (Fig.\ \ref{fig:xps_stobe}c) and measured multilayer data (Fig.\ \ref{fig:growth_cu}c, bottom). A tentative explanation is that the strong interaction of the carbon atoms with the potential of the copper substrate entails a strong chemical environment on the carbon atoms that overwhelms the intramolecular differences and only two sharp main peaks remain.

As mentioned in ref.\ \onlinecite{Diller2013} it was not possible to grow thick multilayers on Cu(111) at RT, i.e., films exceeding approximately two layers (second panel in Fig.\ \ref{fig:growth_cu}). The corresponding NEXAFS and XPS signatures appear to be a mixture of first- and multilayer signals. 
The average tilt angle of the molecules in this bilayer is 25\grad, indicating that for coverages exceeding 1~ML the \mbox{2H-P} molecules start to tilt. From our spectroscopic data solely it cannot be concluded whether all molecules tilt uniformly\cite{Marschall2014} or only the second layer molecules tilt. 
For low deposition temperatures ($T_{evap}$ = 150~K), however, the two layer limit can be overcome and thick multilayer films can be fabricated (Fig.\ \ref{fig:growth_cu}, bottom; see also Figs. S6 and S7). The pronounced dichroism in the NEXAFS data (Fig. \ref{fig:growth_cu}a, bottom) indicates a highly ordered film.  
Fitting of the NEXAFS data (Fig. \ref{fig:growth_cu}b, bottom) yielded an average tilt angle of 30\grad\ with respect to the Cu surface, which may either point to a uniform tilt of all molecules (see illustrations in Fig.\ \ref{fig:growth_cu}) or, as NEXAFS averages over all domains, to a more random distribution. The XPS and NEXAFS spectra of such low temperature grown samples were used in Fig.\ \ref{fig:xps_stobe} and Fig.\ \ref{fig:nexafs_stobe} for the comparison with the simulations, the good agreement points to weakly interacting molecules in the multilayer.

The surprising importance of the second layer for the growth modes of porphine is reminiscent of the case of benzene adsorbed on Cu(111). There, the organic rings also adsorb flat in the first layer.\cite{Xi1994}
Increasing the coverage leads to the formation of a stable bilayer, in which the benzene units of the second layer are oriented perpendicular to the first layer.
 The second layer is special in that sense that it has a slightly higher (5~K) desorption temperature than the molecules in the multilayer.\cite{Xi1994} 
In our case, however, the small NEXAFS angle rules out a completely upright standing second layer. This could be explained by the larger size of the porphine molecule, possibly resulting in a stronger interaction between the first and second layer molecules.

\subsection{Temperature-dependent growth on Ag(111)}
\label{subsec:multi_ag}

To study the influence of the substrate on the growth of \mbox{2H-P} we repeated the experiments on the Ag(111) surface and again investigated the corresponding growth both at RT and LT (Fig.\ \ref{fig:growth_ag}). 
Throughout the whole coverage range from submonolayer to thick multilayers the number and shape of the dominant resonances remain the same (Fig.\ \ref{fig:growth_ag}a) and agree well with the theoretically calculated monomer curves discussed in section \ref{subsec:charac}. 
As mentioned above, in the first layer the molecules adsorb undeformed and parallel to the surface, as indicated by the vanishing 90\grad-curve (Fig.\ \ref{fig:growth_ag}a, top). With increasing coverage the intensity of the 90\grad-curve increases until the dichroism is reversed for the thick multilayer (Fig.\ \ref{fig:growth_ag}a). Because of the coverage-dependent orientation and the related complex damping of the first layer signals, the coverages cannot be quantified accurately, but we estimate the multilayer to be approximately 8 ML thick. 
For the thickest obtained layer the average molecular tilt angle was determined to be 80\grad\ ($\pm$10\grad) i.e., the molecules are oriented almost perpendicular to the surface. Remarkably, this phase differs substantially from the \mbox{2H-P} crystal phase,\cite{Webb1965} which consists of porphine dimers stacked in a T-shape fashion. 
This clearly points to templated growth inducing a multilayer film consisting of uniformly oriented upright standing organic constituents. Templating growth is an important approach to control and optimize the properties of organic thin films.\cite{Fraxedas2002,Witte2004} Accordingly, the epitaxial influence of well-defined substrates on the growth modes of a variety of functional species is intensely investigated.\cite{Ruffieux2002,Koller2006,Klappenberger2011,Lee2014}
A behavior comparable to the one reported here was observed by Söhnchen and co-workers when analyzing the growth of pentacene on Cu(110). They found
flat molecules at low coverages, a perpendicular orientation for thick layers and a third phase in between.\cite{Soehnchen2004} They postulated that the intermediate phase is formed due to the good fit between this structure and the (flat) monolayer, but that for thicker films the stress inside the film becomes larger, which favors a film structure with a lower surface energy.\cite{Soehnchen2004} The same observations and interpretations were reported for pentacene on Au(111).\cite{Beernink2004} We tentatively follow this interpretation as explanation for the different observed phases.

Fig.\ \ref{fig:growth_ag}c shows the corresponding C~1\textit{s} XPS curves, whose shapes change in a systematic manner. Unlike the spectra measured on Cu(111), a very broad structure is visible at (sub)monolayer coverages (Fig.\ \ref{fig:growth_ag}c, top), which transforms to a more defined shape for the multilayer of upstanding molecules. None of the four depicted spectra matches exactly the \mbox{2H-P} monomer spectrum predicted by DFT calculations (Fig.\ \ref{fig:xps_stobe}). The multilayer spectrum shows the highest resemblance, but even with a broadening of 0.7~eV for the simulated curves the sharper experimental features cannot be reproduced exactly. Similarly, DFT simulations of porphine dimers and chains (Fig.\ \ref{fig:dimers}) do not reproduce these observed differences in the measured data, excluding the possible influence of the nearest neighbor molecules. 
 
\begin{figure}[p]
 \includegraphics[width=0.42\columnwidth,clip]{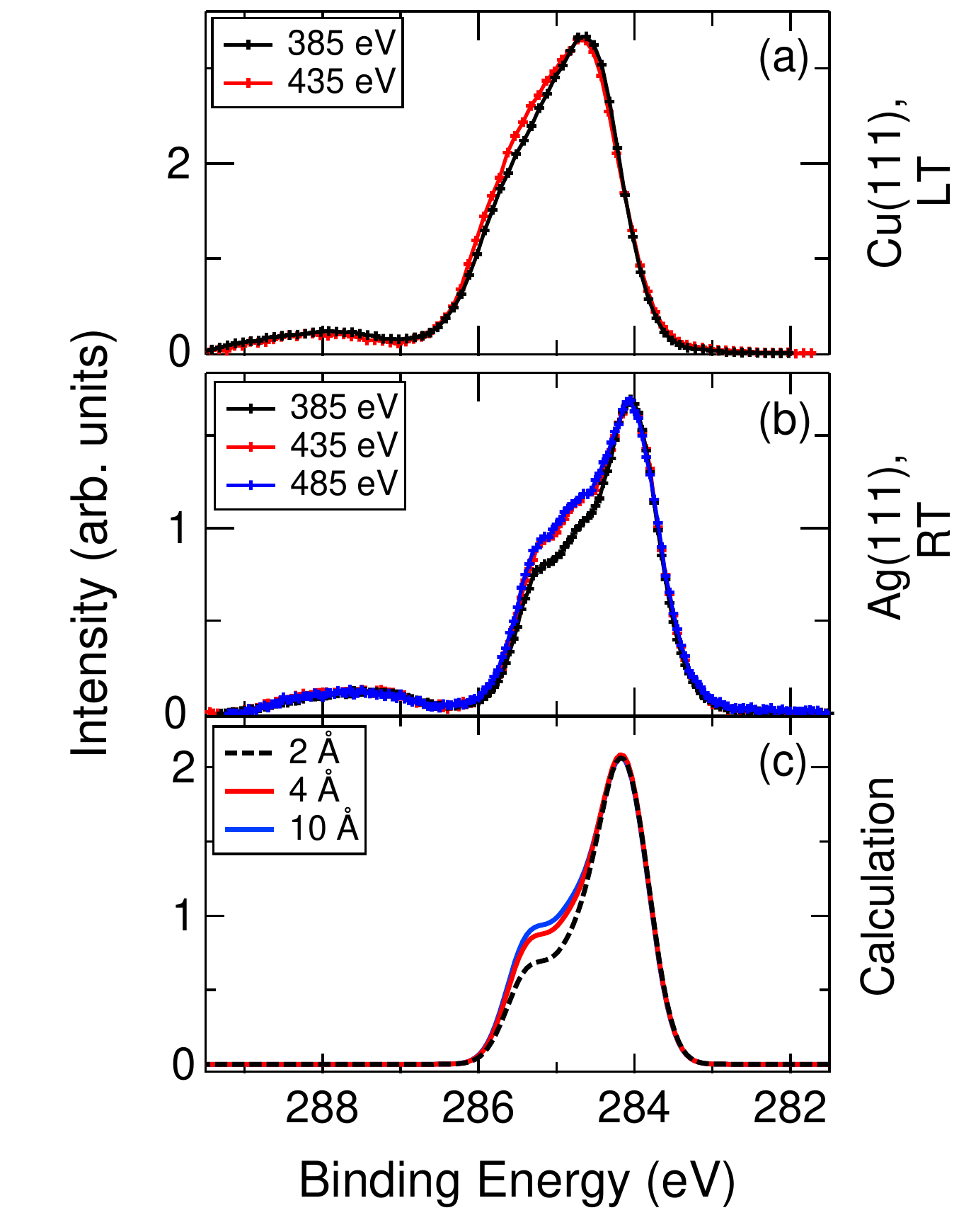}
 \caption{Photon energy dependence of C~1\textit{s} XP spectra of different \mbox{2H-P} multilayers. The spectra are normalized to the peak at lower energies to compensate for the different photon fluxes. (a) Signatures of \mbox{2H-P} deposited at low temperature on Cu(111) agree excellently with the simulation (cf.\ Fig.\ \ref{fig:xps_stobe}) and hardly show any dependence on the photon energies, while those of the highly ordered RT structure at Ag(111) (b) are narrower and photon energy dependent. The latter finding can be explained by (c) simulations for an upright-standing geometry and assuming attenuation of the signal originating from the lower parts of the molecules.}
\label{fig:lambda}
\end{figure}

Dosing \mbox{2H-P} at low temperature on Ag(111) results in the same molecular coverage as dosing at room temperature. However, the average tilt angle of the molecules is 30\grad\ (Fig.\ \ref{fig:growth_ag}, bottom) instead of 80\grad\ for the layers grown at RT and thus identical to the angle measured for a multilayer grown on Cu(111) at LT (Fig.\ \ref{fig:growth_cu}, bottom). The corresponding XP spectrum differs from that of the ordered layer consisting of upright-standing \mbox{2H-P} molecules and is much more similar to both the calculated curve for the monomer and the C~1\textit{s} spectrum of the multilayer grown on Cu(111) at low temperatures.

\subsection{Photon-energy dependence of the photoemission spectra}
The C~1\textit{s} photoemission signal corresponding to the RT multilayer on Ag not only exhibits a modified (sharper) shape, but is additionally shifted to lower binding energies (see Fig.\ \ref{fig:lambda} and Fig.\ S5), which is an indication that the upright room-temperature phase of 2H-P/Ag(111) is much more ordered than the low-temperature films on both Ag(111) and Cu(111). 
Fig.\ \ref{fig:lambda} shows that the C~1\textit{s} spectra of room temperature and low temperature films not only differ in shape, width and energy position, but also react differently to a change in the incidence photon energy $\hbar\omega$. For the flatter LT layer on Cu(111) hardly any change in the relative intensities of the spectroscopic signatures is observed (Fig.\ \ref{fig:lambda}a), whereas the intensity of the high-energy shoulder of the perpendicular 2H-P layers (Fig.\ \ref{fig:lambda}b) clearly depends on the photon energy, i.e., the kinetic energies of the photoelectrons. A variation of the kinetic energy of the photoelectrons leads to a modified inelastic mean free path of the electrons.\cite{Seah1979} Therefore measurements with 385~eV (E$_{kin}$ $\sim$ 100~eV) are much more surface-sensitive and only probe the topmost parts of the molecules. Photoelectron diffraction effects are neglected here, due to the large integration angle selected by the operation mode of the electron energy analyzer. A first tentative explanation of the photon dependence in Fig.\ \ref{fig:lambda}b takes into account that the molecules are oriented differently. For flat molecules the photon energy $\hbar\omega$ should not change the shape of the spectra, as at normal emission electrons originating from inequivalent carbon atoms have to travel the same distance in the film. For perpendicularly oriented molecules the photoemission signal from atoms at the bottom of the porphine are attenuated due to the larger distance traveled through the material before reaching the detector. To quantify this effect, we performed depth-resolved XPS simulations. 
To this end we considered two perpendicular orientations of the molecule where either one of the iminic nitrogen atoms, or one of the pyrrolic nitrogen atoms is closer to the surface. Fig.\ \ref{fig:lambda}c displays the superposition of both. The topmost carbon atoms are set to $z=0$, i.e., their signal is not attenuated. The intensities of the signals originating from all other atoms $i$ at depth $z_i$ are modified according to
\begin{equation}
 I = I_{0}\cdot \exp \left(- \frac{z_i}{\lambda} \right)
 \label{eq:damping}
\end{equation}
where $I_{0}$ is the non-attenuated intensity at $z=$ 0, which is always set to $1$ in the simulation and $\lambda$ is the attenuation length. The resulting shape for different values of $\lambda$ varies for both orientations. Since the porphine molecule is symmetric the effect of the attenuation is not as strong as it would be for an asymmetric molecule, as there are always weak and strong contributions from the same kind of carbon atom. Hence, only at short attenuation lengths are differences predicted in the spectrum. Even though for some systems the inelastic mean free path can be as short as 3~\AA\ at 100~eV kinetic energy,\cite{Seah1979} our derived values for $\lambda$ seem to be smaller than would be expected from literature values of various organic compounds.\cite{Tanuma1994,Lamont1999} This might be explained by the fact that in our approach packing effects, as well as small deviations from an ideal perpendicular orientation, and different orientations of the porphine molecules are not taken into account, which might have a small quantitative effect.
Nevertheless, the trend of the photon energy dependence of the experimental curves is reproduced by the calculations and the results are consistent with the molecular orientation derived from NEXAFS spectroscopy measurements. 


\subsection{Temperature-induced re-orientation}
To get more insight into layer formation pathways, we investigated how the orientation of the molecules responds to changes in the temperature after deposition. To this end, multilayers of \mbox{2H-P} were dosed on cold Ag(111) and Cu(111) substrates, which leads to films with average adsorption angles of 25-30\grad. Slow annealing to room temperature leads to a partial irreversible reorientation of the molecules on both substrates, namely to an increase in tilt angles with respect to the substrate (Fig.\ S6e and i), though the effect is more modest on the Cu(111) substrate.
Unfortunately, the re-orientation is on both substrates accompanied by a partial desorption of the molecules as evidenced by XPS (Fig.\ S6f and j), so that no distinctive new phases could be achieved. A different behavior was observed for a different batch of molecules which contained chlorine and oxygen contaminants. After evaporation at LT again thick films with an average tilt angle of 30-40\grad\ were achieved (Fig.\ S6 c and g). For these films slow annealing to RT led to temperature-stable \mbox{2H-P} multilayers which completely switched to a perpendicular orientation on both substrates without substantial molecule loss (Fig.\ S6 d and h). This implies that the contaminants stabilize the layers (either by a chemical effect or by introducing a different intermolecular spacing) or lower the energy of the transition and, most importantly, that by this procedure ordered, upright standing layers can be achieved not only on Ag(111), but also on Cu(111).


This molecular re-orientation on Cu(111) allows the self-metalation of \mbox{2H-P} on Cu(111)\cite{Diller2013} to be studied.
For bilayers deposited at room temperature and thicker films grown at low temperatures (mostly flat molecular orientation in both cases) all molecules are metalated at 393~K as evidenced by the single peak present in the N 1s XPS (Fig.\ S7a and b). 
However, the multilayer with the chlorine and oxygen contaminations, (grown at LT, then warmed up to RT), which has already switched to an upright orientation, remains nearly unchanged after annealing to 393~K. Even after further annealing to 433~K this contaminated film is not fully metalated. This indicates that (i) the nitrogen atoms need to be close to the surface in order to capture a copper atom from the substrate, which is prevented by the limited mobility of the stacked porphines, and (ii) that, presumably, even the first layer of porphine molecules, which is in direct contact with the substrate, has already begun to tilt.

\begin{figure}[p]
 \includegraphics[width=0.5\textwidth,clip]{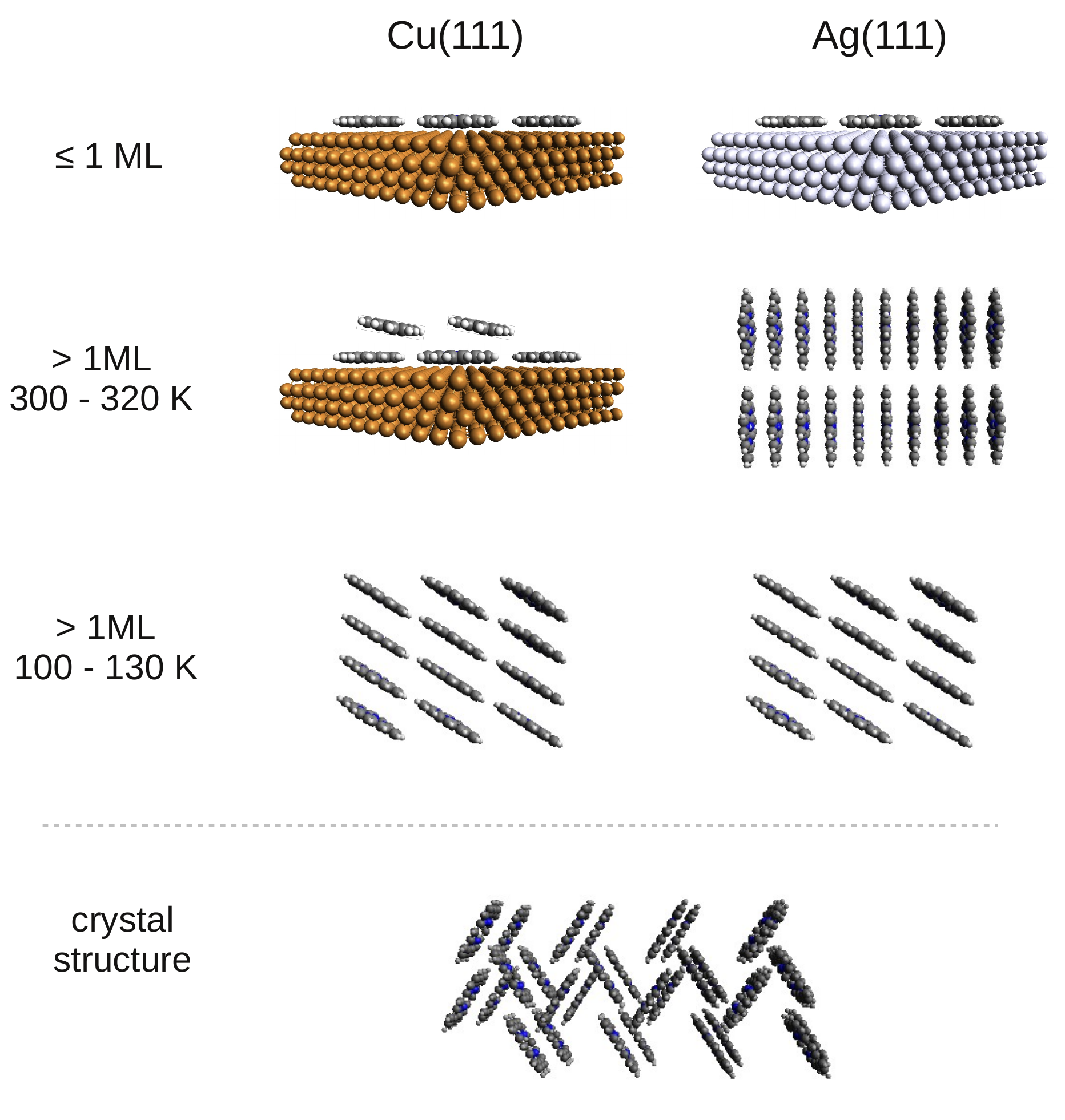}
 \caption{Overview of the temperature-dependent templated growth of porphine thin films on the (111) facets of copper and silver and comparison with crystal structure (taken from ref.\ \onlinecite{Webb1965}). The absence of a support denotes thick multilayers.}
\label{fig:summary}
\end{figure}

\section{Summary}
We explored the growth of a prototypical tetrapyrrole compound, 2H-porphine (2H-P), on the coinage metal surfaces Ag(111) and Cu(111) using a combination of X-ray photoelectron spectroscopy (XPS) and near-edge X-ray absorption fine structure (NEXAFS) spectroscopy and simulated spectra from density functional theory (DFT) calculations. Polarization-dependent NEXAFS measurements reveal the adsorption geometries of the porphines: On both substrates for coverages up to one monolayer the molecules bind without appreciable distortion and parallel to the respective metal surface. For higher coverages the orientation of the molecules depends on the chosen substrate and its temperature during the growth of the films (see Fig.\ \ref{fig:summary}). Multilayers grown at low temperatures (LT) exhibit a similar average tilting angle ($\approx$ 30\grad) on both substrates. 
The corresponding carbon XP and NEXAFS spectra agree very well with the simulated gas phase spectra, and therefore the contributions of inequivalent atoms can be successfully disentangled. 
Specifically, the features observed in both spectra can be decomposed into two groups: peaks at lower binding energies stemming from carbon atoms exclusively connected through C-C bonds, and a second set of peaks at higher binding energies originating from C-N bonded environments.

A remarkably different behavior is observed for molecules dosed at room temperature. On Cu(111) the growth is limited to a coverage of approximately two layers, while on Ag(111) thick multilayers can be grown without restriction. 
The molecules in these multilayers are, in contrast to the 2H-P bulk crystal structure, uniformly oriented with the molecular plane perpendicular to the metal surface. 
Notably, different molecular orientations result in a modified shape of the C~1\textit{s} XPS curves. The dependence of this shape on the incidence photon energy can be rationalized using depth-resolved DFT calculations by taking into account the different attenuation of the various C~1\textit{s} contributions of varying depths. On the other hand, simulations of ionization energies for differently stacked molecules show no indication for a packing-induced modification of the XP spectra.

Interestingly the adsorption geometry also has an influence on the self-metalation on Cu(111), which is prevented or at least hindered for the perpendicularly adsorbed porphines, an effect that was attributed to the reduced contact with the substrate.

A seemingly simple organic-metal surface hybrid system thus presents a surprisingly rich behavior, where the growth conditions and interfacial bonding gives rise to distinct layer geometries, none of which reflects the ordering principles of the bulk organic material.

\section{Acknowledgments}
This work was supported by the European Research Council (ERC Advanced Grant MolArt no. 247299). We thank Klaus Hermann for support and discussions regarding the DFT StoBe calculations. We thank Alexei Nefedov for help during synchrotron experiments and Christof Wöll for kindly providing access to the HE-SGM end-station. The authors acknowledge the Helmholtz-Zentrum Berlin-Electron storage ring BESSY II for provision of synchrotron radiation at beamline HE-SGM. Traveling costs for the BESSY measurements covered by Helmholtz-Zentrum Berlin are gratefully acknowledged. A.C.P. acknowledges a Marie Curie Intra-European Fellowship for Career Development (project no. 274842).

\bibliographystyle{aipnum4-1}
%

\end{document}